\UseRawInputEncoding

\documentclass{ieeeaccess}
\usepackage{cite}
\usepackage{amsmath,amssymb,amsfonts}
\usepackage{graphicx} 
\usepackage{textcomp}

\usepackage{algorithm}
\usepackage{algpseudocode}

\usepackage{listings}

\usepackage{caption}

\usepackage{color}

\definecolor{dkgreen}{rgb}{0,0.6,0}
\definecolor{gray}{rgb}{0.5,0.5,0.5}
\definecolor{mauve}{rgb}{0.58,0,0.82}

\def\BibTeX{{\rm B\kern-.05em{\sc i\kern-.025em b}\kern-.08em
    T\kern-.1667em\lower.7ex\hbox{E}\kern-.125emX}}
\begin{document}
\history{Date of publication xxxx 00, 0000, date of current version xxxx 00, 0000.}
\doi{10.1109/ACCESS.2017.DOI}

\title{Workload Prediction in P4 Programmable Switches}
\author{\uppercase{Boyang Yan}\authorrefmark{1}}
\address[1]{NC State University (e-mail: byan4@ncsu.edu)}



\begin{abstract}
The rapid expansion of cloud services and their unpredictable workload demands present significant challenges in resource management. Traditional resource management approaches, primarily based on static rules and thresholds, often fail to ensure cost-effectiveness and optimal resource utilization. This research introduces a predictive model designed to forecast traffic demand, aiming to shift from a reactive to a proactive resource management approach. By integrating advanced predictive analytics with the capabilities of P4 programmable switches, this study seeks to enhance the efficiency of resource utilization and improve system robustness. The goal is to equip organizations with the agility and economic efficiency required to navigate the complexities of dynamic cloud environments effectively. This approach not only promises to refine microservice resource allocation but also supports the broader objective of fostering more resilient and efficient cloud infrastructures.
\end{abstract}


\titlepgskip=-15pt

\maketitle

\section{Introduction}
\label{sec:introduction}
The rapid proliferation of cloud computing has significantly altered the landscape of software development and deployment, presenting both new opportunities and complex challenges. As enterprises increasingly adopt cloud services to leverage the benefits of scalability, flexibility, and cost-effectiveness, the need for advanced resource management strategies becomes paramount. Traditional approaches, which largely depend on static allocation rules and simplistic threshold-based scaling, are proving inadequate in handling the dynamic and often unpredictable nature of cloud workloads. These conventional methods often lead to either resource over-provisioning, resulting in wasteful expenditure, or under-provisioning, which can compromise service quality and user experience.
In response to these challenges, the field of cloud computing has seen a paradigm shift towards more agile and adaptive resource management techniques. Microservices architecture, a key component of modern cloud applications, exemplifies this shift. It involves decomposing applications into smaller, independently scalable services, each running in its own process and communicating with lightweight mechanisms, such as HTTP resource APIs. This architectural style enhances agility and speeds up deployment but also introduces significant complexity in resource management. Each microservice may scale differently based on its specific workload, requiring a nuanced approach to resource allocation that can anticipate and react to changes in demand in real-time.
Predictive resource scheduling emerges as a crucial strategy in this context. By employing advanced algorithms that can forecast future demands based on historical data, cloud providers can dynamically allocate resources in a manner that optimizes cost and performance. This research focuses on developing a predictive model that not only addresses the inefficiencies of traditional resource management approaches but also leverages the unique capabilities of P4 programmable switches. These switches allow for highly flexible and programmable network behavior, which is critical in managing the high-speed data flows typical of modern data centers.
The advent of smart scheduling techniques, which incorporate real-time data and predictive analytics, represents a significant advancement in cloud resource management. These techniques enable data centers to operate more efficiently, reduce operational costs, and improve the overall quality of service by minimizing latency and avoiding service disruptions. Our study is situated at this innovative juncture, aiming to harness the potential of P4 programmable switches to revolutionize resource scheduling in cloud environments. By integrating smart scheduling with predictive analytics, this research contributes to the ongoing evolution of cloud computing, pushing the boundaries of what is possible in dynamic and complex cloud environments.

\subsection{\label{sec:questions}Research Questions}
Predictive workload management and resource scheduling are increasingly critical in cloud computing, particularly as demand for cloud services grows and the architecture of applications becomes more complex. The literature on this subject is rich with various approaches aimed at enhancing the efficiency and responsiveness of cloud resources through predictive analytics and intelligent scheduling.

\subsubsection{Historical Context and Evolution}
Initially, research in this field focused on static resource allocation strategies that did not account for the fluctuating nature of cloud workloads. Early works by Calheiros et al. (2011) and Beloglazov et al. (2012) laid the groundwork by demonstrating the limitations of static resource allocation and the potential benefits of adopting dynamic resource provisioning based on real-time data analysis \cite{buyya2009modeling, beloglazov2010energy}. These studies underscored the need for more adaptive resource management techniques that could respond to changes in workload demand without human intervention.

\subsubsection{Advances in Predictive Modeling}
As the field evolved, more sophisticated predictive models were introduced. Techniques such as time series analysis, machine learning, and neural networks began to be applied to forecast future demands in cloud environments. For instance, Lorido-Botrán et al. (2014) explored the use of autoregressive integrated moving average (ARIMA) models for predicting the workload in cloud systems, paving the way for more advanced predictive algorithms\cite{lorido2012auto}.
The introduction of machine learning techniques for workload prediction has been particularly transformative. Researchers like Tordsson et al. (2012) began employing various forms of regression analysis, support vector machines, and neural networks to predict the workload and thus optimize resource allocation\cite{tordsson2012cloud}. These methods have proven effective in not only forecasting demand but also in reducing the operational costs associated with under or over-provisioning of resources.

\subsubsection{Role of P4 Programmable Switches}
The advent of P4 programmable switches has introduced new possibilities in network management and resource scheduling. The programmability of these switches allows for the implementation of complex algorithms directly into the network fabric, enabling real-time, adaptive control of data flows and resource allocation. This capability is particularly pertinent in high-traffic data center environments where traditional networking hardware struggles to keep pace with dynamic and high-speed requirements.
Studies by Bosshart et al. (2014) and Kim et al. (2015) have explored the applications of P4 programmable switches in network traffic management and security \cite{bosshart2014p4, kim2015band}. More recently, researchers have begun investigating their potential in facilitating smart scheduling and predictive resource management in cloud computing infrastructures. The flexibility of P4 switches to adapt to changing network conditions and their ability to process data packets with custom-defined rules makes them ideal for implementing predictive scheduling algorithms that require rapid adjustments to network behavior.

\subsubsection{Emerging Trends and Current Research}
The latest research in this area is focusing on integrating artificial intelligence (AI) with P4 programmable technologies to further enhance predictive capabilities. AI-driven approaches are being explored to handle even more complex scenarios involving multiple variables and highly dynamic environments. These studies aim to leverage the inherent strengths of P4 programmable switches to execute AI algorithms efficiently, ensuring that resource scheduling can be as responsive and proactive as possible. There are five open questions in this area. This project will only focus on scheduling for predictions of the VMs/Micro-services’ expected resource utilization. In another ward, prediction dynamic workload demands.
\begin{itemize}
  \item Smart scheduling. Before selecting servers to run a set of new VMs/Micro-services, the scheduler can contact the Real-time Controller for predictions of the VMs/Micro-services’ expected resource utilization. With this information, the scheduler can select servers to balance the disk IOPS load or to reduce the likelihood of physical resource exhaustion in oversubscribed servers.
  \item Smart cluster selection. Before selecting a cluster in which to create a VM/Micro-services deployment, the cluster selection system can query the Real-time Controller for a prediction of maximum deployment size. With this information, this system can select a cluster that will likely have enough resources.
  \item Smart power oversubscription and capping. During a power emergency (when the power draw is about to exceed a circuit breaker limit), the power capping system can query the Real-time Controller for predictions of VM/Micro-services workload interactivity, before portioning the available power budget across servers. Ideally, VMs/Micro-services executing interactive workloads should receive all the power they may want, to the detriment of VMs/Micro-services running batch and background tasks. Alternatively, the VM/Micro-services scheduler can request interactivity predictions before selecting servers, so that interactive and delay-insensitive workloads are segregated in different sets of servers.
  \item Scheduling server maintenance. When a server starts to misbehave, the health monitoring system can query the Real-time Controller for the expected lifetime of the VMs/Micro-services running on the server. It can thus determine when maintenance can be scheduled, and whether VMs/Micro-services need to be live-migrated to enable maintenance without unavailability. In addition, the VM/Micro-services scheduler can use the lifetime predictions to co-locate VMs/Micro-services that are likely to terminate roughly at the same time. This could facilitate other types of maintenance, such as OS updates.
  \item Recommending VM/Micro-services and deployment sizes. The cloud platform could provide a service to its customers that recommends the appropriate VM/Micro-services size and number of VMs/Micro-services at the time of each deployment. Using the Real-time Controller predictions of workload class and resource utilization, the service could recommend deployments where VMs/Micro-services predicted to be delay-insensitive would be more tightly sized than interactive VMs/Micro-services.
\end{itemize}

\subsection{\label{sec:contributions}Contributions}
\begin{itemize}
    \item Data center networks require high reliability and high-speed traffic (100Gbps), so our algorithms must be interpretable and cannot be black-box.
    \item We modify the decision tree (DT) to the Binary Decision Tree (BDT). Compared with the DT, our BDT:
    \begin{itemize}
        \item Supports faster training.
        \item Generates fewer rules.
        \item Satisfies switch constraints better.
    \end{itemize}
    \item Moreover, as learning algorithms (particularly deep learning) have shown their superior performance, some more complicated learning models are emerging for networking. However, their:
    \begin{itemize}
        \item High computational complexity and
        \item Large storage requirement
    \end{itemize}
    are causing challenges in the deployment on switches.
\end{itemize}

\subsection{\label{sec:structure}Structure}
This paper is organized as follows. We begin with a 'Background' section, which provides a comprehensive review of related work and a general overview of the field. This is followed by the 'Algorithms' section, where we describe the computational models and predictive algorithms employed in our research. Next, the 'Methods' section details the experimental setup and data preprocessing techniques. The 'Results' section presents the outcomes of our experiments, including data analysis and findings. Subsequent to this, the 'Discussion' section interprets the results, exploring the implications and limitations of our study. Finally, the 'Conclusion' section summarizes the key points of our research, offering insights and potential directions for future work.

\section{\label{sec:introduction}Background}
In this section, we explore the architecture of microservices, followed by an examination of the P4 switch and its constraints. We then delve into a comparative analysis of eBPF and P4 switch technologies, highlighting the distinct features and operational differences between the two. This comparison is crucial for understanding the specific advantages that P4 switches offer in network management and resource scheduling within cloud computing environments. Through this discussion, we aim to clarify the suitability of each technology in various scenarios and their implications for modern network architectures.
\subsection{Microservices architecture}
Microservices architecture diverges from the traditional monolithic design paradigm, where all functional components of an application are interwoven into a single indivisible unit\cite{Bmic15}. By contrast, microservices encapsulate discrete functionalities into independently deployable modules, facilitating modular development, deployment, and scaling. This architectural refinement aligns well with cloud computing platforms that excel in offering elastic computing resources, thereby enabling services to be scaled in or out based on real-time demand.

The crux of managing a microservices-based application lies in the effective orchestration of service replicas to ensure seamless scalability and resilience. Horizontal scaling, or the dynamic adjustment of service instances, emerges as a cornerstone strategy in this context. It ensures that each microservice can independently scale according to its workload, thereby optimizing resource allocation and minimizing latency. 

Nonetheless, the stochastic nature of web traffic and service demands poses a significant challenge, often resulting in either resource over-provisioning — leading to unnecessary expenditure — or under-provisioning, which risks breaching service level agreements (SLAs) and deteriorating user experience.

Cloud computing complements the microservices architecture by providing a scalable and flexible execution environment that supports rapid provisioning and de-provisioning of resources in response to varying workloads. The synergy between microservices and cloud platforms encapsulates the potential for achieving high levels of operational efficiency, agility, and cost-effectiveness. Nevertheless, the ephemeral and volatile nature of microservices workloads necessitates a predictive approach to scaling that transcends reactive and heuristic based methods. Predictive scaling, underpinned by analytical and machine learning models, offers a prescient mechanism to forecast demand and proactively adjust resource allocations, thereby ensuring optimal service performance and resource utilization.

This research explores and validates the efficacy of predictive workload scaling within the microservices architecture, aiming to refine resource management strategies, enhance QoS, and reduce operational costs. Through the development and empirical assessment of predictive models, this study seeks to contribute to the burgeoning field of cloud-native application optimization, presenting a pragmatic pathway to harnessing the full potential of microservices and cloud computing technologies.

\subsection{P4 Switch and Its Constraints}
A P4 switch refers to a network switch that is programmable using the P4 language. P4, which stands for "Programming Protocol-independent Packet Processors," is a high-level programming language designed specifically for networking applications. It enables network engineers and researchers to define how packets are processed by the data plane of a network switch, router, or network interface card (NIC), independent of the underlying hardware. 

The flexibility provided by P4 programmability allows for the customization of network behaviors, optimization of performance, and rapid prototyping of new network features that are traditionally fixed in network devices. This adaptability is crucial in modern network environments, which require rapid changes in network policies and functions to support diverse applications and services.

P4 switches are often used in data centers, research facilities, and by network equipment vendors who require the ability to experiment with and deploy new network functionalities efficiently.

\subsubsection{P4 Switch data plane}
The data plane of a P4 switch incorporates a Protocol Independent Switch Architecture (PISA). Upon receipt of a packet, it undergoes an initial transformation into a Packet Header Vector (PHV) via the parsing mechanism. Subsequently, the PHV traverses through a pipeline structured with multiple stages of match-action units (MAUs). Within this pipeline, if a specific header field within the PHV, such as the destination port, aligns with an entry in a predefined table, it activates corresponding processes within the action unit linked to that entry. The PHV, after undergoing these designated processes, is then restructured back into a packet format by the deparser. PISA enables the configuration of custom processes, including defining tables and associated actions in the P4 language, which are then implemented within the MAUs.

\subsubsection{Matching Constraints}
For P4 language, the standard library \texttt{core.p4} defines three standard match kinds:
\begin{enumerate}
    \item \textbf{Exact match:} The key must match exactly with the field in the rule.
    \item \textbf{Ternary match:} The key is ANDed with a mask, then compared with the value for exact match.
    \item \textbf{Longest prefix match (LPM):} This guarantees that the mask is a series of consecutive bits 1 followed by a series of consecutive bits 0. It is widely supported by diverse devices.
\end{enumerate}
Other libraries (e.g., \texttt{v1model.p4}) may define additional match kinds such as range match, fuzzy match, but these are not available on many hardware targets.

\subsubsection{Memory Constraints}
Each stage is equipped with two high-speed types of memory:
\begin{itemize}
    \item \textbf{TCAM:} Ternary Content-Addressable Memory suitable for fast table lookups.
    \item \textbf{SRAM:} Static Random Access Memory is used to store exact match table entries and stateful registers.
\end{itemize}
Packet processing in Match Action Units (MAUs) is limited to simple instructions like integer additions and bit shifts. The entire pipeline should occupy as few stages as possible to prevent packet delays or stalls.

\subsection{eBPF (Extended Berkeley Packet Filter) and P4 comparison}
eBPF is another advanced network programming mechanism that operates within the Linux kernel, allowing for the dynamic modification and enhancement of network behavior without recompiling the kernel. It is primarily used for high-performance packet processing, network monitoring, and security applications. eBPF programs can be written in a high-level language like C and compiled into eBPF bytecode, which runs in a virtual machine-like environment inside the kernel, offering extensive flexibility and safety \cite{vieira2020fast}. Unlike eBPF, P4 is specifically designed for data plane programming, allowing developers to specify how devices process packets. This can be deployed in various platforms, including those supporting eBPF, to tailor the packet processing pipeline specifically to the needs of a network, often on programmable hardware such as ASICs and FPGAs. A notable implementation is the NIKSS, which leverages P4 for programming software-defined networking (SDN) data paths, and translates P4 programs to run in the eBPF execution environment, combining the strengths of both P4 and eBPF \cite{osinski2022novel}. As a result, P4 is more widely in use, we will only be focused on P4 in this project.

\section{Machine Learning Experiments Management}
Nowadays, machine learning workflow involves large sets of runs (a.k.a iterations) over different machine learning assets and quickly becomes complex \cite{idowu2022asset}. There are lots of challenges to management in Machine Learning Experiments. The challenges encountered during machine learning experiments are often related to the lack of explicit tooling support to address experiment management concerns, including reproducibility, replicability, traceability, explainability, interpretability, collaboration, and auditability.
\subsection{MLFlow}
In this research, I used MLFlow, which is an open-source platform designed to manage the end-to-end machine learning lifecycle. It provides tools to help data scientists and ML engineers manage the stages of the ML process, including experimentation, reproducibility, and deployment. All of my experiments trace, and the reproduction package is managed by MLFlow. 

\section{Hyperparameter Search}
Hyperparameter search is a critical component of machine learning, impacting model performance significantly. Various methods exist for tuning hyperparameters, each with its advantages and limitations.

\subsection{Random Search}
Random search involves generating and evaluating random inputs to the objective function. This method is efficient because it does not make any assumptions about the structure of the objective function \cite{arden2022hyperparameter}. Random search is particularly beneficial when domain knowledge might bias the optimization process, enabling the discovery of non-intuitive solutions. Additionally, it can be the most effective approach for complex problems with noisy or discontinuous regions in the search space, where algorithms that depend on reliable gradients might struggle.

\subsection{Grid Search}
Grid search is a method for selecting the best combination of algorithms and hyperparameters by exhaustively testing and validating each possible combination. The goal is to identify the combination that delivers the highest performance, which can then be used as the chosen predictive model. While grid search necessitates high-dimensional spaces with defined boundaries, it is often easily parallelizable because the evaluations of different hyperparameter values are independent of one another. While comprehensive, it can be computationally expensive and inefficient for large spaces \cite{arden2022hyperparameter}.

\subsection{Bayesian Optimization Method}
Unlike random search and grid search, Bayesian Optimization efficiently navigates the parameter space to identify the configuration that best optimizes an overall evaluation criterion (OEC), such as accuracy, AUC, or likelihood. This method achieves a balance between exploration and exploitation, allowing it to home in on optimal configurations more quickly. By leveraging prior experiment results to inform decisions on which hyperparameter sets to evaluate next, Bayesian optimization can identify the best hyperparameters in less time.

\subsection{Genetic algorithm using TPOT}
Genetic algorithms mimic the process of natural selection, where only the species best adapted to environmental changes survive, reproduce, and pass on their genes to the next generation. The time required to run these algorithms can be substantial, depending on the size of the search space. However, the number of generations and the population size can be adjusted. Each individual in the search space represents a potential solution to a specific problem.

TPOT automates the machine learning process by intelligently exploring and evaluating various machine learning configurations and pipelines. It selects only those that deliver the best results. TPOT iterates through different algorithms to identify the optimal one, choosing the candidate with the highest accuracy score. Additionally, it can combine two or more algorithms to create a hybrid algorithm, further enhancing performance.

\subsection{SHERPA}
SHERPA is a hyperparameter optimization library for machine learning algorithms. It is designed to identify the best hyperparameter values, particularly for models that require expensive iterative function evaluations, such as deep neural networks. SHERPA provides users with multiple interchangeable hyperparameter optimization strategies, each of which can be advantageous at different stages of algorithm development.

\subsection{Optuna}
Optuna is a software framework for automating the hyperparameter tuning process for efficiency and flexibility\cite{akiba2019optuna}. It automatically identifies optimal hyperparameter values using several samplers, including grid search, random, Bayesian, and evolutionary algorithms. Optuna categorizes its optimization techniques under two distinct strategies: sampling and pruning. Algorithms determine the optimal parameter combination by focusing on regions where hyperparameters produce superior outcomes. There are two sampling methods: relational sampling, which leverages relationships between parameters, and independent sampling, which samples each parameter separately. It provides a define-by-run API, making it easy to dynamically construct the search space. Optuna supports pruning of unpromising trials based on intermediate results, which can significantly reduce computational resources \cite{shimazoemethod}. Optuna can be integrated with MLflow to log and manage hyperparameter optimization experiments. This combination leverages MLflow's robust tracking and Optuna's optimization capabilities.

\section{Hyperparameter performance comparison}

\section{Algorithms}
\subsection{\label{sec:baseline}Workload Prediction ( baseline)}
There are five types of methods for Time Series Prediction, which are the Regression Method, Multiple Regression Method, Exponential Smoothing, Stochastic Forecasting Technique, Soft Computing Based Forecasting Technique\cite{mahalakshmi2016survey}.

\subsubsection{Vector Auto Regression (VAR) \cite{holden1995vector}}
Vector Auto Regression (VAR) is a statistical model used to capture the linear interdependencies among multiple time series. VAR models generalize the univariate autoregressive model by allowing for multivariate time series. It's an extension of the autoregressive (AR) model to multiple time series, which is why it's called 'vector' autoregression. VAR models are widely used for analyzing the dynamic impact of random disturbances on the system of variables. After estimating a VAR model, impulse response functions can be used to analyze the effect of a one-time shock to one of the innovations on current and future values of the endogenous variables. While VAR models are very flexible and can be used in a variety of contexts, they have limitations. They can require a large amount of data to estimate accurately and can suffer from the "curse of dimensionality" when dealing with a large number of time series.
A VAR model is formulated as follows for \( k \) time series and a VAR of order \( p \):

\[
Y_{i,t} = c_i + \sum_{j=1}^{p} \sum_{k=1}^{K} A_{ij} Y_{k,t-j} + \varepsilon_{i,t}
\]

where \( Y_{i,t} \) is the value of the \( i \)-th variable at time \( t \), \( c_i \) is a constant, \( A_{ij} \) are the coefficients of the lagged values of the \( j \)-th time series, and \( \varepsilon_{i,t} \) is the error term.

\subsubsection{Support vector regression (SVR)}
Support Vector Machines (SVMs) are a robust set of supervised learning methods used for classification, regression, and outliers detection \cite{vapnik1974theory}. The versatility of SVMs lies in their ability to handle both linear and non-linear data effectively, using a set of mathematical functions known as kernels. The traditional application of SVMs is in binary classification, but they can be extended to solve regression problems (known as Support Vector Regression). Support vector regression (SVR), introduced by Drucker et al. \cite{drucker1996support}, is an extension of the well-known SVMs and was devised for single-output regression. SVR uses a quadratic program (QP) to obtain the predictions of a single output. SVR is particularly well-suited for situations where the prediction of continuous values is required, and the data may exhibit complex patterns that linear models cannot capture. SVR maintains all the main features that characterize the algorithm for classification (SVM): It uses the same principles of minimization of an error function, maximizing the margin, and controlling the model complexity using a regularization parameter. Similar to SVMs, SVR can perform linear regression in the original input space or it may map the input into high-dimensional feature spaces through the use of kernel functions. The kernel trick is particularly useful when the relationship between the input variables and the target variable is non-linear.

SVR offers several advantages for predictive modeling:
\begin{itemize}
    \item \textbf{Robustness}: SVR is robust to outliers and is capable of handling non-inear relationships effectively. This makes it well-suited for real-world data which often contains noise and anomalies.
    \item \textbf{Flexibility}: Through the use of different kernels, such as linear, polynomial, and radial basis functions, SVR can model complex relationships without the need for extensive data transformation. This flexibility allows it to adapt to various types of data characteristics.
    \item \textbf{Regularization}: The regularization parameter in SVR provides a means to control overfitting. This parameter helps in maintaining a balance between achieving a low error on the training data and minimizing the model complexity, which ensures good generalization to unseen data.
\end{itemize}

\subsubsection{Random Forest Regressor}
The Random Forest Regressor is an ensemble learning technique for regression tasks, which builds upon the concept of bagging (bootstrap aggregating) and the decision tree algorithms \cite{liaw2002classification}. It creates a forest of trees where each tree is slightly different from the others, and their collective output is used to make final predictions. This method is widely used due to its robustness, simplicity, and effectiveness across a wide range of datasets.

Random Forest Regressor offers several significant advantages:

\begin{enumerate}
    \item \textbf{Bootstrap Sampling}: Each decision tree is trained on a random sample of the data, drawn with replacement from the original dataset, which may lead to some observations being repeated in each sample.
    \item \textbf{Feature Randomness}: When constructing trees, the split at each node is made not from the best split among all features but from a random subset of features. This increases diversity among the trees and is crucial for reducing correlation between them.
    \item \textbf{Tree Construction}: Trees are grown to their maximum length without pruning (fully grown and untrimmed) to capture complex patterns, although this can lead to overfitting.
    \item \textbf{Aggregation for Prediction}: For regression tasks, the final prediction is the average of the predictions from all trees. This averaging reduces noise and generally improves the model's ability to generalize.
\end{enumerate}

\subsubsection{Gradient Boosting Machines (GBM)}
Gradient Boosting Machines (GBM) are a powerful and widely-used ensemble learning technique that builds predictive models from an ensemble of weak learners, typically decision trees. The core idea behind GBM is to iteratively improve the prediction accuracy by focusing on the errors made by previous models.

\subsubsection{Large Bayesian vector auto regressions \cite{banbura2010large}}
Large Bayesian vector auto regressions (BVARs) are an advancement of the traditional vector auto regression (VAR) models, incorporating Bayesian statistical methods to handle systems with a large number of time series. BVARs address the parameter proliferation problem in large VARs by using shrinkage priors. This Bayesian approach helps in dealing with over-parameterization that often accompanies the inclusion of many variables in a model. Despite their complexity, large BVARs with shrinkage can produce credible impulse response functions, which are essential for interpreting the dynamic relationships among variables.

\subsubsection{Explainable Boosted Linear Regression (EBLR)\cite{ilic2021explainable}}
An iterative method that starts with a base model and explains the model's errors through regression trees. It incorporates nonlinear features by residuals explanation and extends to probabilistic forecasting through generating prediction intervals based on the empirical error distribution.

\subsubsection{Time series forecasting using distribution enhanced linear regression\cite{ristanoski2013time}}
This method explores enhancing the performance of linear regression for time series forecasting by employing a grouping-based quadratic mean loss function. The approach involves segmenting the input time series into groups and optimizing both the average loss of each group and the variance of the loss between groups over the entire series. This method aims to create a linear model that not only has low overall error but also exhibits robustness to outliers and sensitivity to distribution changes in the time series.

\section{Methods}
\subsection{Dataset Overview}
This section offers a succinct overview of the datasets employed in this research. We utilize seven publicly available datasets, categorized into four distinct types: VM Traces, Serverless Traces, Microservice Traces, and Traffic Request Traces. Each category has been selected to address different aspects of workload prediction and resource scheduling in cloud environments, providing a comprehensive data foundation for our analysis.

\begin{table}[h] 
\centering 
\caption{Overview of Datasets used in the Project} 
\label{tab:dataset_overview} 
\begin{tabular}{llll}
\hline
\textbf{Dataset Name} & \textbf{Type} & \textbf{Year} & \textbf{Cite} \\ \hline
Azure\cite{cortez2017resource}        & VM Traces              & 2017 & 630 \\
Azure Functions\cite{shahrad2020serverless}  & Serverless Traces      & 2020 & 475 \\
Alibaba AMTrace\cite{wang2022characterizing} & Microservice Traces    & 2022 & 7   \\
Google Borg\cite{gao2020task} & Microservice Traces    & 2020 & 326 \\
Calgary-HTTP\cite{ebadifard2021autonomic} & Traffic request Traces & 1994 & 54  \\
ClarkNet-HTTP\cite{ebadifard2021autonomic}         & Traffic request Traces & 1995 & 54  \\
NASA-HTTP\cite{ebadifard2021autonomic}             & Traffic request Traces & 1995 & 54  \\
Saskatchewan-HTTP\cite{ebadifard2021autonomic}     & Traffic request Traces & 1995 & 54  \\ \hline
\end{tabular}
\end{table}

\subsubsection{Traffic request Traces}
The datasets are as follows: Calgary-HTTP, ClarkNet-HTTP, NASA-HTTP, and Saskatchewan-HTTP. Each dataset consists of the following five variables:

\begin{itemize}
    \item \textbf{Host making the request:} Hosts are identified as either local or remote.
    \item \textbf{Timestamp:} In the format "DAY MON DD HH:MM:SS YYYY", where DAY is the day of the week, MON is the name of the month, DD is the day of the month, HH:MM:SS is the time of day using a 24-hour clock, and YYYY is the year.
    \item \textbf{Filename} of the requested item.
    \item \textbf{HTTP reply code.}
    \item \textbf{Bytes} in the reply.
\end{itemize}

\paragraph{Calgary-HTTP}
The Calgary-HTTP dataset comprises a comprehensive record of HTTP requests made to the Department of Computer Science's WWW server at the University of Calgary, located in Calgary, Alberta, Canada. The dataset encompasses a period of approximately one year, from October 24, 1994, to October 11, 1995, totaling 353 days. It includes 726,739 HTTP requests, with timestamps recorded at a resolution of one second. This dataset provides a significant resource for analyzing web server traffic patterns over an extended period.

\paragraph{ClarkNet-HTTP}
The ClarkNet-HTTP dataset includes detailed records of all HTTP requests made to the ClarkNet WWW server over a two-week period. ClarkNet is a comprehensive Internet service provider serving the Metro Baltimore-Washington DC area. The data collection was conducted in two phases: the first from 00:00:00 on August 28, 1995, to 23:59:59 on September 3, 1995, and the second from 00:00:00 on September 4, 1995, to 23:59:59 on September 10, 1995. Each phase lasted exactly seven days, culminating in a total of 3,328,587 HTTP requests captured over the 14-day span. All timestamps in the dataset are recorded with a one-second resolution, facilitating precise analysis of request timing and frequency.

\paragraph{NASA-HTTP}
The NASA-HTTP dataset encapsulates two months of comprehensive HTTP request data from the NASA Kennedy Space Center WWW server in Florida. Data collection was methodically executed in two consecutive periods: the first from 00:00:00 on July 1, 1995, to 23:59:59 on July 31, 1995, spanning 31 days, followed by the second from 00:00:00 on August 1, 1995, to 23:59:59 on August 31, 1995, which inadvertently states seven days but correctly accounts for 31 days. Over this two-month duration, a total of 3,461,612 HTTP requests were logged. All timestamps in the dataset are recorded with one-second resolution, providing detailed temporal data for analysis of web server traffic patterns.

\paragraph{Saskatchewan-HTTP}
The Saskatchewan-HTTP dataset comprises an extensive collection of HTTP requests spanning seven months to the University of Saskatchewan's WWW server, located in Saskatoon, Saskatchewan, Canada. The data were meticulously collected from 00:00:00 on June 1, 1995, through 23:59:59 on December 31, 1995, covering a total of 214 days. During this period, the server processed 2,408,625 HTTP requests. All timestamps in the dataset are captured with a resolution of one second, providing detailed insights into the server's request handling over the extended duration. This dataset serves as a valuable resource for analyzing web traffic dynamics and server load distribution.

\paragraph{Data Mutation}
Data mutation is a fundamental aspect of data wrangling and preprocessing. It involves combining and summarizing datasets into a continuous time series, capturing metrics such as the number of packets and the total packet size.

\subsection{time series decomposition / Sampling Techniques}
Sampling techniques are used to select a subset of data from a larger dataset. The goal is often to make inferences about the entire dataset or to reduce the size of the dataset for analysis. There are a couple of ways to work on Sampling.
\begin{itemize}
    \item \textbf{Random Sampling}: Each member of the population has an equal chance of being selected.
    \item \textbf{Stratified Sampling}: The population is divided into strata, and samples are taken from each stratum.
    \item \textbf{Systematic Sampling}: Every nth member of the population is selected.
    \item \textbf{Cluster Sampling}: The population is divided into clusters, and entire clusters are randomly selected.
    \item \textbf{Convenience Sampling}: Samples are taken from a group that is conveniently accessible.
\end{itemize}
In this research, first I use Stratified Sampling to divide the whole dataset into homogeneous subpopulations (Day, Week, Month) called strata (the plural of stratum). Then, sampled using Simple Random Sampling, and chose one-day, one-week, and one-month data. Finally, use Time Series Decomposition, a technique used to break down a time series into its constituent components. These components typically include:
\begin{itemize}
    \item \textbf{Observed data}: Observed data
    \item \textbf{Trend Component}: This captures the long-term progression of the series (e.g., the increasing or decreasing direction over time).
    \item \textbf{Seasonal Component}: This captures the repeating short-term cycle (e.g., daily, weekly, monthly, or yearly patterns).
    \item \textbf{Residual (or Irregular) Component}: This captures the random variation or noise left after removing the trend and seasonal components.
\end{itemize}

\begin{figure}[H]
    \centering
    \includegraphics[width=0.45\textwidth]{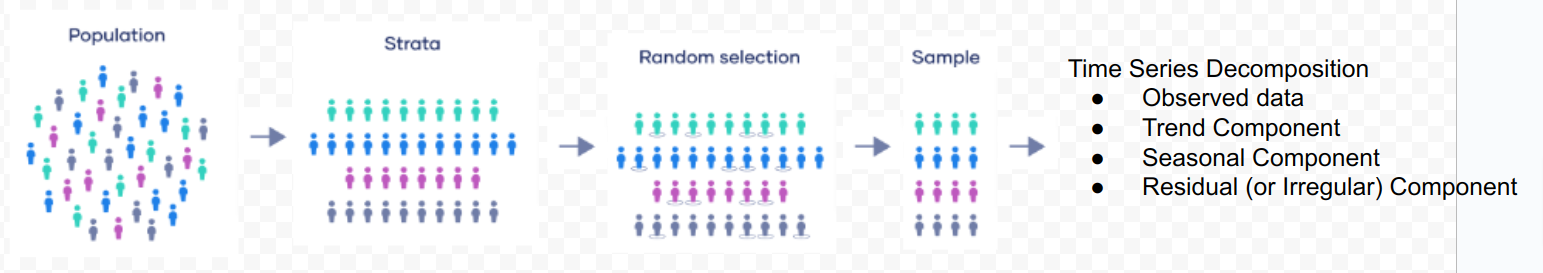}
    \caption{time series decomposition workflow}
    \label{fig:day-numOfPacketDecomposition}
\end{figure}

The day dataset period in hour. The week dataset period in day (24 hours). The month dataset period also in day (24 hours).

Time Series Decomposition result below.
\subsubsection{day(24 hours) number of Packet Decomposition}

\begin{figure}[H]
    \centering
    \includegraphics[width=0.45\textwidth]{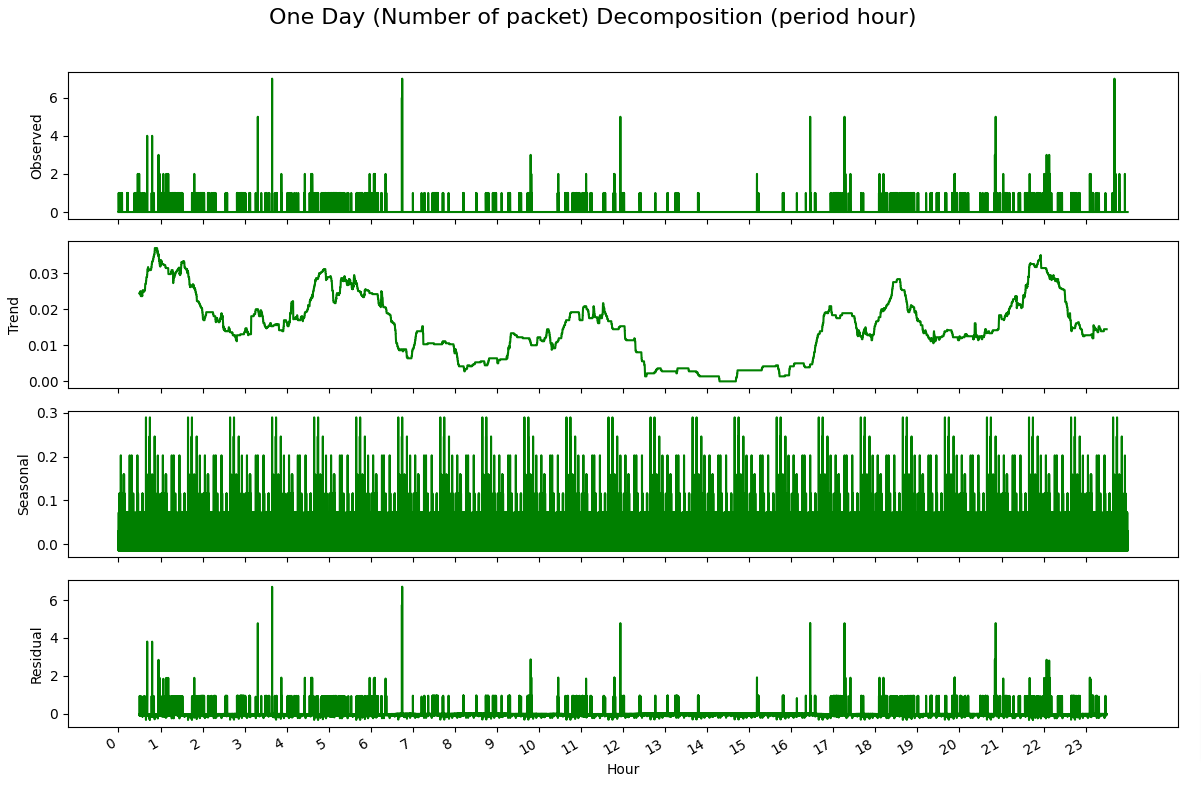}
    \caption{day(24 hours) number of Packet Decomposition}
    \label{fig:day-numOfPacketDecomposition}
\end{figure}

\subsubsection{day (24 hours) sum of Packet Size Decomposition}

\begin{figure}[H]
    \centering
    \includegraphics[width=0.45\textwidth]{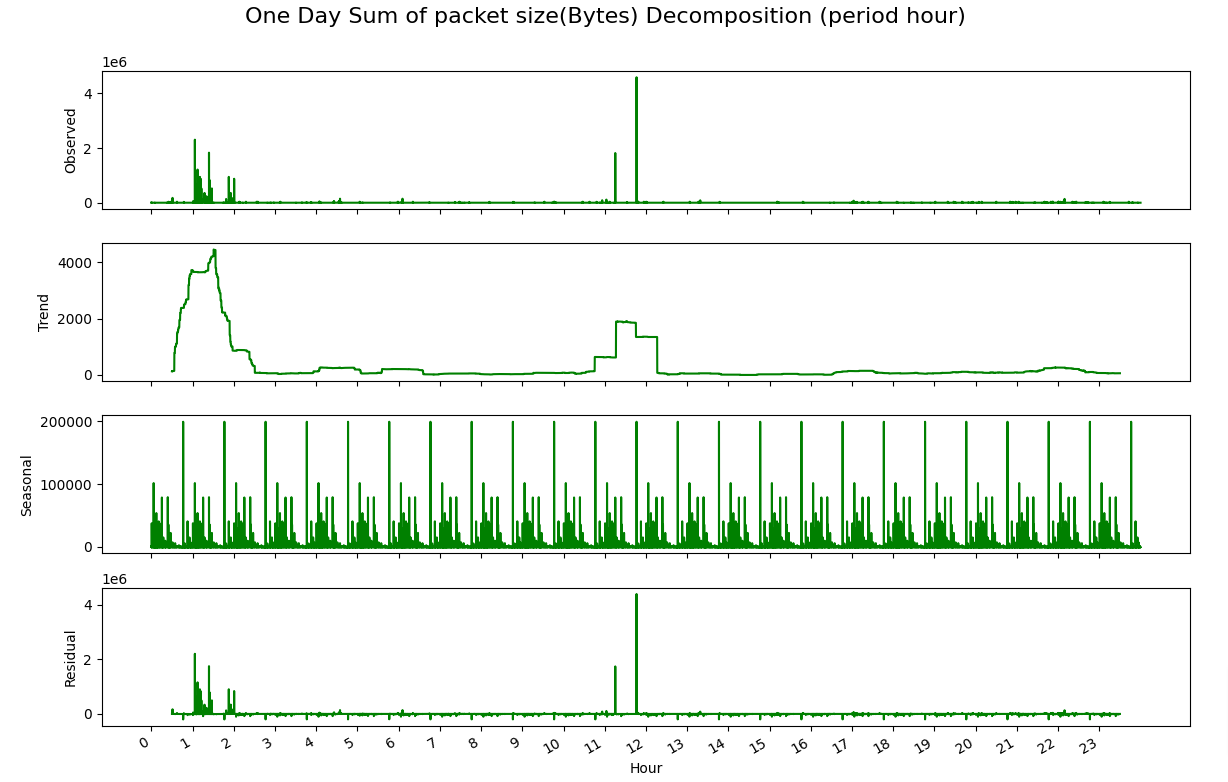}
    \caption{day (24 hours) sum of Packet Size Decomposition}
    \label{fig:day-sizeDecomposition}
\end{figure}

\subsubsection{One week (number of Packet) Decomposition (period day)}

\begin{figure}[H]
    \centering
    \includegraphics[width=0.45\textwidth]{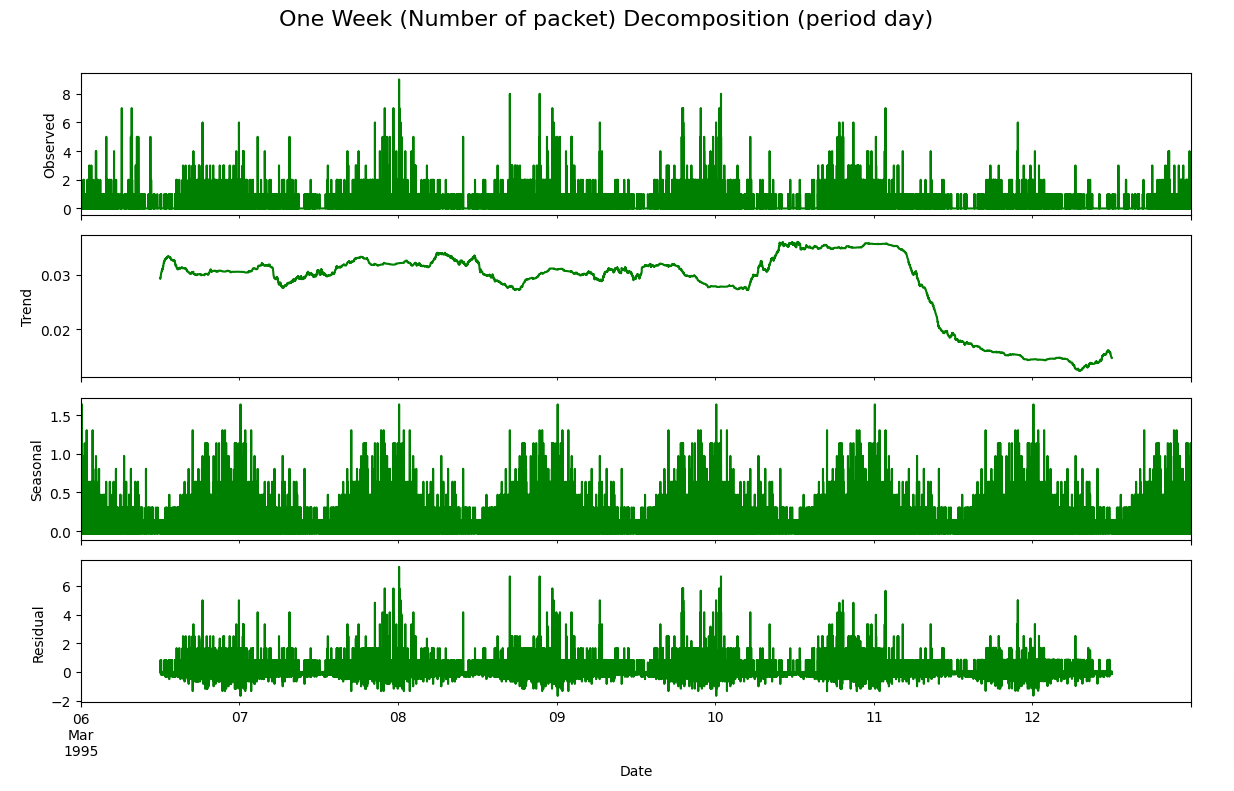}
    \caption{One week (number of Packet) Decomposition (period day)}
    \label{fig:week-numOfPacketDecomposition}
\end{figure}

\subsubsection{One week sum of Packet Size Decomposition}

\begin{figure}[H]
    \centering
    \includegraphics[width=0.45\textwidth]{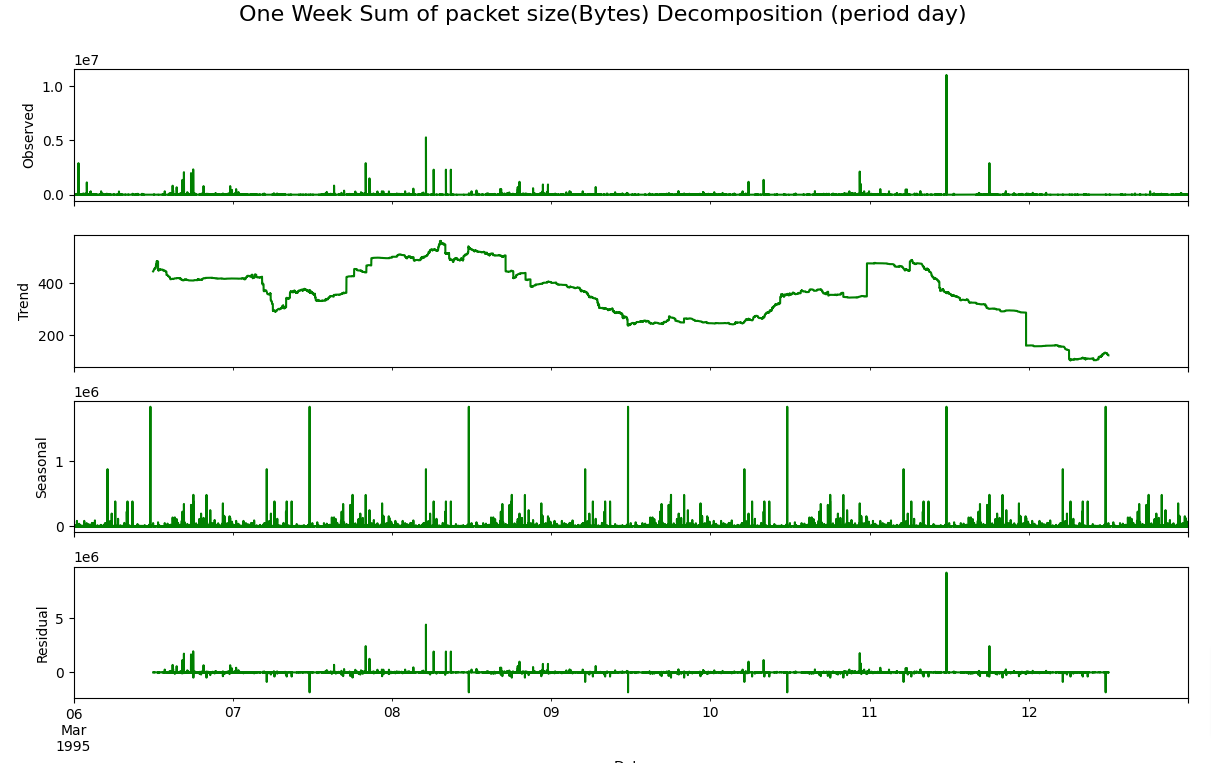}
    \caption{One week sum of Packet Size Decomposition}
    \label{fig:week-packetSizeDecomposition}
\end{figure}

\subsubsection{One Month (Number of packets) Decomposition (period day)}

\begin{figure}[H]
    \centering
    \includegraphics[width=0.45\textwidth]{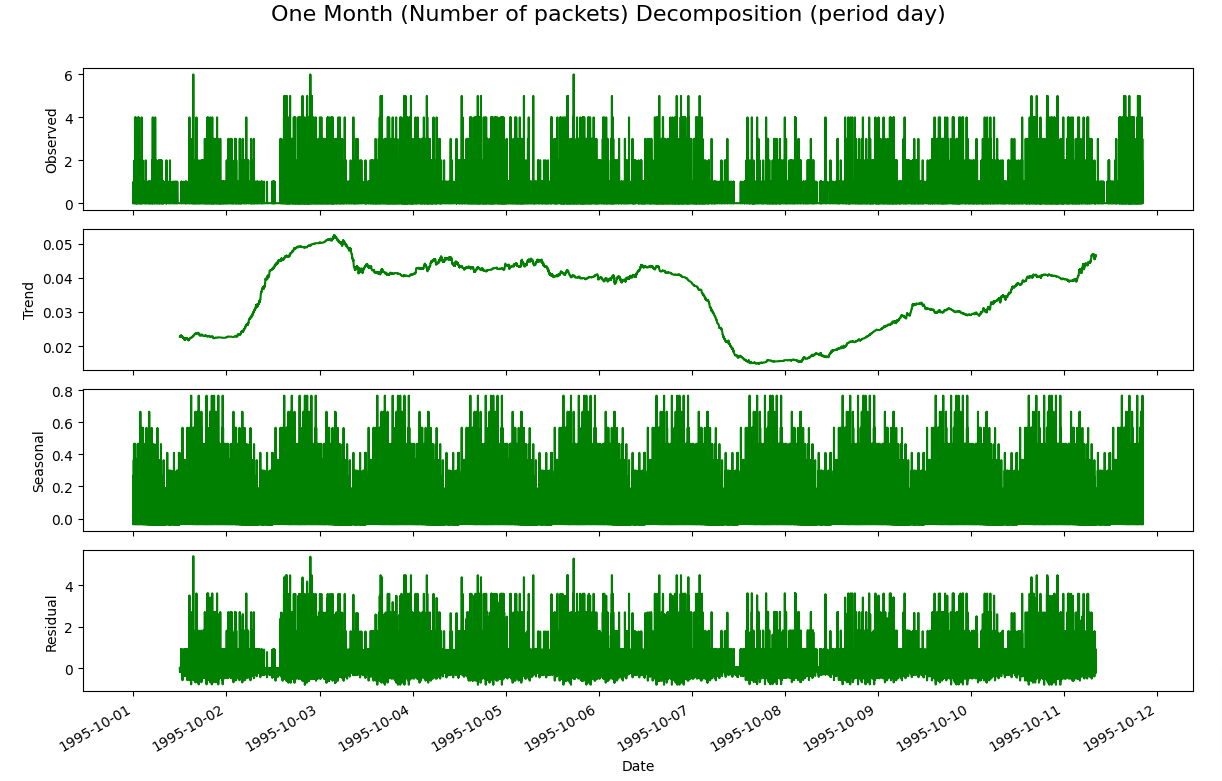}
    \caption{One Month (Number of packets) Decomposition (period day)}
    \label{fig:month-numOfPacketsDecomposition}
\end{figure}

\subsubsection{One Month Sum of packet size(Bytes) Decomposition (period day)}

\begin{figure}[H]
    \centering
    \includegraphics[width=0.45\textwidth]{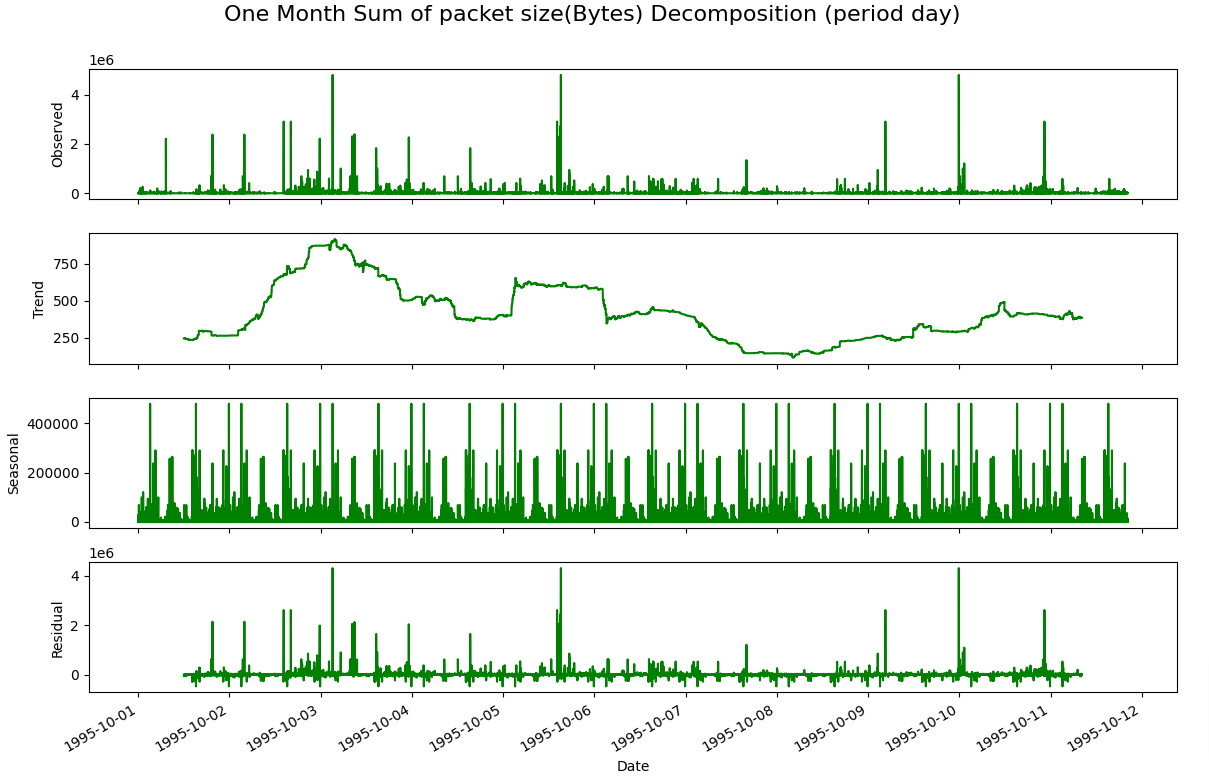}
    \caption{One Month Sum of packet size(Bytes) Decomposition (period day)}
    \label{fig:month-packetSizeDecomposition}
\end{figure}

All datasets, including the number of packets and the sum of packet sizes, exhibit seasonality and contain significant levels of irregular components. Also, there are trends in all of the datasets. The number of packets shows more noise compared to the packet sizes. Therefore, I will first assess the stationarity of the observed data and then apply Holt-Winters exponential smoothing (triple exponential smoothing) to the observed data.

\subsection{Check for Stationarity (Augmented Dickey-Fuller (ADF) test)}
It is imperative that all time series variables within the dataset exhibit stationarity, a statistical property characterized by a constant mean and variance over time \cite{lu2013stationarity}. The Augmented Dickey-Fuller (ADF) test is a prevalent method for assessing stationarity. The ADF test operates under the null hypothesis that posits the time series as non-stationary. A p-value from the ADF test that is less than the predetermined significance level leads to the rejection of the null hypothesis, thereby affirming the stationarity of the time series. Should the data prove to be non-stationary, one recommended approach is to apply differencing to the time series data to achieve stationarity.

\subsection{Data Smoothing}
Smoothing techniques are kinds of data preprocessing techniques to remove noise from a data set. This allows important patterns to stand out.
\subsubsection{Moving average (MA) smoothing}
It is a simple and common type of smoothing used in time series analysis and forecasting. Here time series derived from the average of last kth elements of the series.

The formula provided is:
\[
S_t = \frac{X_{t-k} + X_{t-k+1} + X_{t-k+2} + \ldots + X_t}{k}
\]

\begin{itemize}
  \item $S_t$ is the smoothed value at time $t$.
  \item $X_t$ is the actual value at time $t$.
  \item $k$ is the number of periods over which the average is calculated.
\end{itemize}

\subsubsection{Exponential smoothing}
Exponential smoothing is a weighted moving average technique. In the moving average smoothing the past observations are weighted equally, In this case smoothing is done by assigning exponentially decreasing weights to the past observations.

\[
S_0 = X_0
\]

\[
S_t = \alpha \cdot X_t + (1 - \alpha) \cdot S_{t-1} \quad \{ t > 0, \, 0 < \alpha < 1 \}
\]

In the above equation, we can see that $(1 - \alpha)$ is multiplied by the previously expected value $S_{t-1}$ which is derived using the same formula. This makes the expression recursive, and if you were to write it all out on paper, you would quickly see that $(1 - \alpha)$ is multiplied by itself again and again. And this is why this method is called exponential.

\subsubsection{Double exponential smoothing}
Single Smoothing does not excel in the data when there is a trend. This situation can be improved by the introduction of a second equation with a second constant β.

t is suitable to model the time series with the trend but without seasonality.

\[
S_0 = X_0
\]

\[
B_0 = X_1 - X_0
\]

\[
S_t = \alpha \cdot X_t + (1 - \alpha) \cdot (S_{t-1} + B_{t-1})
\]

\[
B_t = \beta \cdot (S_t - S_{t-1}) + (1 - \beta) \cdot B_{t-1}
\]

\[
\alpha, \beta \in (0, 1)
\]

Here it is seen that $\alpha$ is used for smoothing the level and $\beta$ is used for smoothing the trend.

\subsubsection{Triple exponential smoothing}
It is also called as Holt-winters exponential smoothing. it is used to handle the time series data containing a seasonal component.
double smoothing will not work in case of data contain seasonality.so that for smoothing the seasonality a third equation is introduced.

\[
S_0, F_0 = X_0
\]

\[
B_0 = \frac{\sum_{i=0}^{L-1} (X_{L+i} - X_i)}{L^2}
\]

\[
S_t = \alpha \cdot (X_t - C_{\%L}) + (1 - \alpha) \cdot (S_{t-1} + \phi \cdot B_{t-1})
\]

\[
B_t = \beta \cdot (S_t - S_{t-1}) + (1 - \beta) \cdot \phi \cdot B_{t-1}
\]

\[
C_{\%L} = \gamma \cdot (X_t - S_t) + (1 - \gamma) \cdot C_{\%L}
\]

\[
F_{t+m} = S_t + B_t \cdot \sum_{i=1}^{m} \phi^i + C_{\%L}
\]

\[
\alpha, \beta, \gamma \in (0, 1)
\]

In the above, $\phi$ is the damping constant. $\alpha$, $\beta$, and $\gamma$ must be estimated in such a way that the MSE (Mean Square Error) of the error is minimized. The best $\alpha$, $\beta$, and $\gamma$ will determine by Optuna hyperparameter tuning process.

\subsection{\label{sec:solution}Proposed Solution}
Recursive Random Projection (RRP) emerges as a robust technique for regressing high-dimensional data, drawing from foundational works \cite{bingham2001random, fern2003random, dasgupta2008random}. RRP is fundamentally a Nyström algorithm, primarily employed in approximating the eigenvectors and eigenvalues of matrices \cite{platt2005fastmap}. Comparable to Principal Components Analysis (PCA) in its full implementation, Nyström's approach can be executed more rapidly in its variant forms, such as FASTMAP \cite{faloutsos1995fastmap}. While RRP's applicability extends to various tasks, our study focuses exclusively on its utility in regression analyses, denoted herein as RRP Regression.
\begin{itemize}
  \item Localization: take current example, walk it down the cluster tree to find its relevant leaf
  \item Anomaly detection: Anomalous if you are far away from everything else in your relevant leaf
  \item Classification: Localization + report mode class label in relevant leaf
  \item Regression: Localization + report median class label in relevant leaf
\end{itemize}

The pseudocode below illustrates the procedures involved in the Recursive Random Projection (RRP) Regression.

\begin{lstlisting}
Algorithm: RandomProjectionRegression
Input: data, stop_depth, target_index
Output: predictions

Class RandomProjectionRegression:
    Initialize(data, stop_depth, target_index):
        self.data ← data
        self.stop_depth ← stop_depth
        self.target_index ← target_index
        self.tree ← BuildTree(data.to_numpy(), stop_depth)

    Function BuildTree(candidates, enough):
        if not isinstance(candidates, numpy_array):
            candidates ← convert_to_numpy_array(candidates)

        if length_of(candidates) ≤ enough:
            median_value ← calculate_median(candidates[:, target_index])
            return {"type": "leaf", "median": median_value}

        east_pivot, west_pivot, east_items, west_items ← Split(candidates)
        return {
            "type": "node",
            "east_pivot": east_pivot,
            "west_pivot": west_pivot,
            "east_child": BuildTree(east_items, enough),
            "west_child": BuildTree(west_items, enough)
        }

    Function Split(candidates):
        pivot ← random_choice(candidates)
        east_pivot ← FindFarest(pivot, candidates)
        west_pivot ← FindFarest(east_pivot, candidates)
        distance_c ← CalculateDistance(east_pivot, west_pivot)

        if distance_c = 0:
            mid_point ← length_of(candidates) // 2
            return east_pivot, west_pivot, candidates[:mid_point], candidates[mid_point:]

        all_distance ← []
        for candidate in candidates:
            distance_a ← CalculateProjectionDistance(candidate, west_pivot, east_pivot, distance_c)
            append (distance_a, candidate) to all_distance

        sort all_distance by first element of each tuple
        mid_point ← length_of(all_distance) // 2
        east_items ← first_half_of_sorted(all_distance)
        west_items ← second_half_of_sorted(all_distance)
        return east_pivot, west_pivot, east_items, west_items

    Function Predict():
        predictions ← []
        for row in data:
            prediction ← LocalizeAndPredict(convert_row_to_numpy(row), tree)
            append prediction to predictions
        return predictions

    Function LocalizeAndPredict(data_point, node):
        if node["type"] = "leaf":
            return node["median"]
        if CalculateDistance(data_point, node["east_pivot"]) < CalculateDistance(data_point, node["west_pivot"]):
            return LocalizeAndPredict(data_point, node["east_child"])
        else:
            return LocalizeAndPredict(data_point, node["west_child"])

    Static Function FindFarest(pivot, candidates):
        max_distance ← 0
        farest_point ← pivot
        for candidate in candidates:
            distance ← CalculateDistance(pivot, candidate)
            if distance > max_distance:
                max_distance ← distance
                farest_point ← candidate
        return farest_point

    Static Function CalculateDistance(p1, p2):
        return square_root(sum((v1 - v2)² for each (v1, v2) in zip(p1, p2)))

    Static Function CalculateProjectionDistance(candidate, west_pivot, east_pivot, c):
        distance_a ← CalculateDistance(candidate, west_pivot)
        distance_b ← CalculateDistance(candidate, east_pivot)
        return (distance_a² + c² - distance_b²) / (2 * c)

\end{lstlisting}

\subsection{Framework Overview}
Our whole in-network intelligence framework is
shown in Fig.  For our packet prediction task. we obtain a trained BDT. Then, we encode the prediction rules of the BDT into ternary match table entries, and finally install these entries on the P4 program of the switch. Now, for an
arriving packet on the switch, the whole prediction task is model-free, and the P4 program will look up the packet’s header fields and find its matched table entry and the related class label.
\begin{figure}[H]
    \centering
    \includegraphics[width=0.5\textwidth]{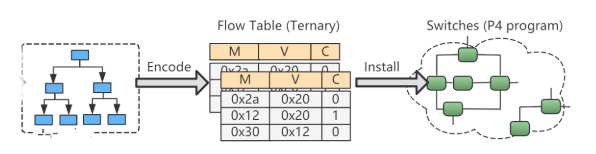}
    \caption{Framework Overview}
    \label{fig:p4}
\end{figure}

\subsection{\label{sec:evaluation}Evaluation Metrics}
In this research, we mainly use two evaluation metrics: mean absolute error (MAE) and model size. Since most of the algorithms are stochastic, Scott-Knott is the statistical analysis we used in our study. All of the details are below.

\subsubsection{Overfitting}

\subsubsection{Mean Absolute Error}
To assess the accuracy of the time series models used in our forecasting, we compute the Mean Absolute Error (MAE). MAE has been widely favored by researchers across various fields as a robust metric for evaluating errors in time series analysis \cite{krishna2018connection, willmott2005advantages, willmott2006use}. It measures the difference between two continuous variables, represented as paired observations \(X\) and \(Y\), which correspond to the same underlying phenomenon. Common examples include comparing predicted values versus observed values, assessing changes over time, and comparing different measurement techniques. A scatter plot of \(n\) points, where each point \(i\) has coordinates \((x_i, y_i)\), facilitates this comparison. The MAE is calculated as the average distance between each point and the identity line \(Y = X\), both vertically and horizontally:

\begin{equation}
    MAE = \sum_{i=1}^{N} p_i |\hat{y}_i - y_i| = \sum_{i=1}^{N} p_i |e_i|
\end{equation}

Here, \(N\) represents the total number of observations, \(p_i\) denotes the probability of occurrence for each unique value, \(y_i\) is the actual value, and \(\hat{y}_i\) is the predicted value. Lower values of MAE indicate higher accuracy.

While some researchers \cite{amin2013approach} advocate for using other metrics like Mean Squared Error (MSE) or Mean Magnitude of Relative Error (MMRE) for evaluating forecast quality, we choose to focus on MAE due to its clarity and straightforward interpretation. This choice is further justified because:
\begin{itemize}
  \item MAE directly represents the average absolute discrepancy between the forecasted \(\hat{Y}_i\) and actual \(Y_i\) values, providing a clear and intuitive measure of prediction accuracy. This clarity of interpretation is highlighted and preferred in several studies \cite{willmott2005advantages, willmott2006use}.
  \item MMRE can distort error assessment, especially when actual values are zero or close to zero, leading to disproportionately large error values that may misrepresent the model's accuracy.
\end{itemize}

Our focus on MAE supports a transparent and reliable approach to error measurement, mitigating common pitfalls associated with other metrics.

\subsubsection{Root Mean Squared Error (RMSE)}
The Root Mean Squared Error (RMSE) is another widely used metric for measuring the accuracy of predictions or forecasts. RMSE quantifies the average magnitude of errors between the predicted or forecasted values and the actual values, giving more weight to larger errors. It’s an important metric, especially when larger errors are of greater concern.

\subsubsection{Mean Absolute Percentage Error (MAPE)}
The Mean Absolute Percentage Error (MAPE) is a metric used to measure the accuracy of predictions or forecasts as a percentage of the actual values. It quantifies the average percentage difference between the predicted or forecasted values and the actual values.

\subsubsection{Symmetric Mean Absolute Percentage Error (SMAPE)}
The Symmetric Mean Absolute Percentage Error (SMAPE) is a metric used for measuring the accuracy of predictions or forecasts in time series analysis. It’s particularly useful when you want to assess forecast accuracy while considering both overestimation and underestimation errors. SMAPE calculates the percentage difference between predicted or forecasted values and actual values, but it symmetrically treats overestimation and underestimation errors.

\subsubsection{Median Absolute Percentage Error (MDAPE)}
MDAPE stands for Median Absolute Percentage Error. It is a performance metric used to evaluate the accuracy of forecasts in time series analysis. MDAPE is similar to the Mean Absolute Percentage Error (MAPE), but instead of taking the mean of the absolute percentage errors, it takes the median. This makes MDAPE less sensitive to outliers than MAPE.

\subsubsection{Geometric Mean Relative Absolute Error (GMRAE)}
The Geometric Mean Relative Absolute Error (GMRAE) is a metric used to assess the accuracy of predictions or forecasts in time series analysis. It takes the geometric mean of the relative absolute errors, providing a single aggregated measure of forecast accuracy. GMRAE is particularly useful when you want to consider both overestimation and underestimation errors.

\subsubsection{Model Size Consideration}
In this research, the algorithms are designed to operate on networking devices, which typically have limited memory and computational resources. Consequently, the model size becomes a critical metric for evaluating the feasibility and performance of our approach \cite{xie2022mousika}. It is imperative that our algorithms are optimized to have a minimal memory footprint, ensuring that they can be effectively deployed within the constraints of networking hardware. We anticipate that our refined algorithms will achieve a reduced model size, enhancing their suitability for real-time applications on network devices.

Here are a few common ways to determine the size of a machine learning model:

\begin{enumerate}
  \item \textbf{File Size on Disk:} After training a model, you can save it to a file using serialization libraries or built-in functions in your machine learning framework. The size of this file on disk is a direct indication of the model's size. For example, in Python using libraries like TensorFlow or PyTorch, you might save a model and then check the file size using file system commands or Python functions like \texttt{os.path.getsize()}.
  
  \item \textbf{Model Parameters:} The size can also be estimated based on the number of parameters in the model. This includes weights, biases, and other trainable parameters. For deep learning models, frameworks like TensorFlow and PyTorch provide functions to summarize the model, which includes total parameters. Multiply the total number of parameters by the number of bytes per parameter (usually 4 bytes for a floating-point number) to get an approximate size.
  
  \item \textbf{In-Memory Size:} When loaded into memory, the size of the model might be different due to additional overhead or optimizations. You can measure this by monitoring memory usage before and after loading the model, using tools specific to your operating system or runtime environment.
\end{enumerate}

\subsection{Statistical Analysis}
Since most of the algorithms are stochastic, and the different train test splits may result in different samples generated by those algorithms, we run each algorithm 10 times and use statistical analysis to compare results through 10 repeats. Scott-Knott is the statistical analysis we used in our study.

More specifically, Scott-Knott recursively partitions the list of candidates \( l \) into two sub-lists \( l_1 \) and \( l_2 \). The split should maximize the expected mean value before and after the division \cite{tu2020better, xia2018hyperparameter, tu2021mining}:

\begin{equation}
E(\Delta) = \frac{|l_1| * | \bar{l}_1 - \bar{l}| + |l_2| * |\bar{l}_2 - \bar{l}|}{|l|}
\end{equation}

We check if two sub-lists differ significantly by using the Cliff's Delta procedure. The delta value is

\begin{equation}
\Delta = \frac{(\#(x > y) - \#(x < y))}{(|l_1| * |l_2|)}
\end{equation}

for \( \forall x \in l_1 \), and \( \forall y \in l_2 \). Cliff's delta estimates the probability that a value in the sub-list \( l_1 \) is greater than a value in the sub-list \( l_2 \), minus the reverse probability \cite{macbeth2011cliff}. Two sub-lists differ significantly if the delta is not a ``small'' effect (\( \Delta \geq 0.147 \)) \cite{hess2004robust}.

The reason we choose Scott-Knott because (a) it is fully non-parametric and (b) it reduces the number of potential errors in the statistical analysis since it only requires at most \( O(\log_2(N)) \) statistical tests for the \( O(N^2) \) analysis. Other researchers also advocate for the use of this test \cite{gates1978illustration} since it overcomes a common limitation of alternative multiple-comparison statistical tests (e.g., the Friedman test \cite{friedman1937use}) where treatments are assigned to multiple groups (making it hard for an experimenter to distinguish the real groups where the means should belong \cite{carmer1985pairwise}).

\section{Results}
This study evaluated the performance of various predictive modeling techniques for workload prediction in cloud environments using P4 programmable switches. The primary evaluation metric was the Mean Absolute Error (MAE), which directly measures predictive accuracy.

The MAE for each predictive model is summarized in the table below:

\begin{table}[ht]
\centering
\resizebox{0.5\textwidth}{!}{
\begin{tabular}{|l|c|c|}
\hline
\textbf{Algorithm} & \textbf{Number of packets (MAE)} & \textbf{Sum of packets aize (MAE)} \\
\hline
Vector Auto Regression (VAR) & 0.02 & 144.79 \\
\hline
Explainable Boosting Regressor (EBR) & 0.00027974940691152903 & 2.828485179096606e-13 \\ 
\hline
Boosted Linear Regression & 9.016438269161553e-20 & 1.0861660310258403e-21 \\
\hline
Support Vector Regression (SVR) & 0.000306 & 0.000191 \\
\hline
Random Forest Regressor & 0.0000 & 0.0000 \\
\hline
Gradient Boosting Machines (GBM) & 2.845676637322534e-41 & 7.939844244132699e-15 \\
\hline
Recursive Random Projection Regression & 8,089.668 & 8,144.964 \\
\hline
\end{tabular}
}
\caption{Comparison of predictive models based on Mean Absolute Error (MAE).}
\label{tab:mae_comparison}
\end{table}

\pagebreak
\subsection{Gradient Boosting Machines (GBM)}
Best hyperparameters for Count model: \{'n\_estimators': 286, 'max\_depth': 6, 'learning\_rate': 0.27473835134535474, 'min\_samples\_split': 10, 'min\_samples\_leaf': 2\}

Best hyperparameters for Bytes model: \{'n\_estimators': 268, 'max\_depth': 18, 'learning\_rate': 0.16498161952272483, 'min\_samples\_split': 8, 'min\_samples\_leaf': 3\}

\subsection{Random Forest Regressor}
\{'n\_estimators': 105, 'max\_depth': 5, 'min\_samples\_split': 5, 'min\_samples\_leaf': 3\}

\subsection{Support Vector Regression (SVR)}
Bytes: \{'C': 1.6175496437504917, 'epsilon': 0.013824522502100901, 'kernel': 'sigmoid'\}

Count: \{'C': 0.26222410264606155, 'epsilon': 0.017495969009300106, 'kernel': 'linear'\}

\subsection{Boosted Linear Regression}
Bytes (Target1): \{'n\_estimators': 458, 'learning\_rate': 0.2639472208589661, 'max\_depth': 10\}

Count (Target2): \{'n\_estimators': 352, 'learning\_rate': 0.2640148453302873, 'max\_depth': 2\}

\subsection{Explainable Boosting Regressor (EBR)}
Bytes (Target1): \{'learning\_rate': 0.09992124060291752, 'max\_bins': 442, 'max\_interaction\_bins': 21, 'outer\_bags': 16\}

Count (Target2): \{'learning\_rate': 0.07852101013489443, 'max\_bins': 500, 'max\_interaction\_bins': 20, 'outer\_bags': 4\}

\subsection{Vector Auto Regression (VAR)}
\{'maxlags': 7, 'ic': 'hqic'\}

\subsubsection{MAE}
\paragraph{number of Packet}
\begin{figure}[H]
    \centering
    \includegraphics[width=0.45\textwidth]{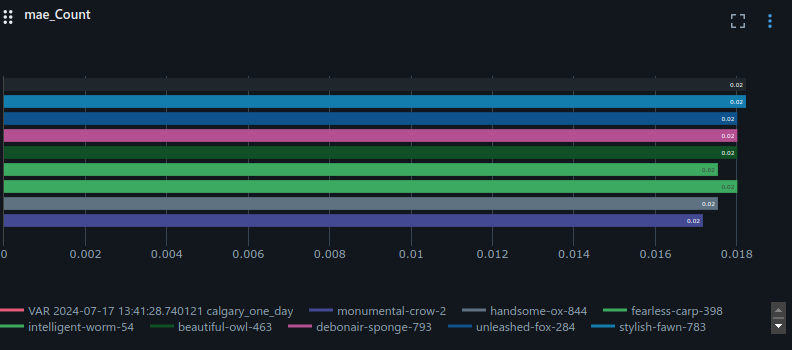}
    \caption{number of Packet}
    \label{fig:week-packetSizeDecomposition}
\end{figure}

\paragraph{sum of Packet Size}
\begin{figure}[H]
    \centering
    \includegraphics[width=0.45\textwidth]{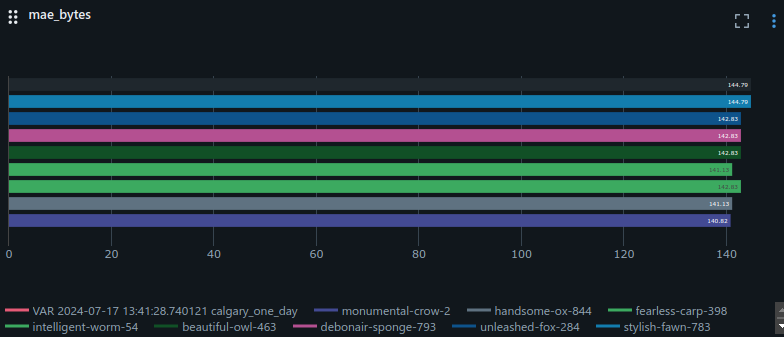}
    \caption{sum of Packet Size}
    \label{fig:week-packetSizeDecomposition}
\end{figure}

\subsubsection{rmse}
\paragraph{number of Packet}
\begin{figure}[H]
    \centering
    \includegraphics[width=0.45\textwidth]{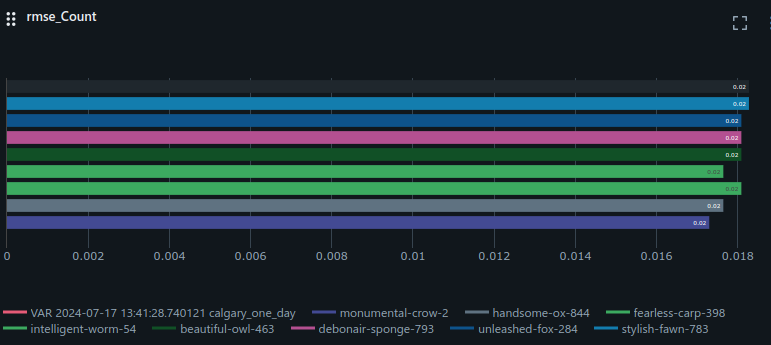}
    \caption{number of Packet}
    \label{fig:week-packetSizeDecomposition}
\end{figure}

\paragraph{sum of Packet Size}
\begin{figure}[H]
    \centering
    \includegraphics[width=0.45\textwidth]{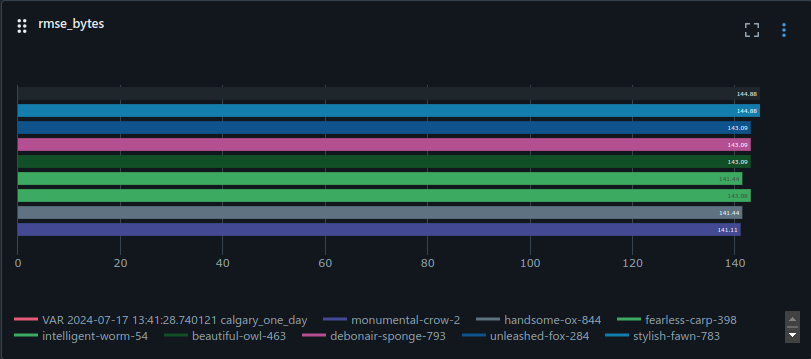}
    \caption{number of Packet}
    \label{fig:week-packetSizeDecomposition}
\end{figure}

\subsection{Recursive Random Projection Regression}
\subsubsection{MAE}
\paragraph{number of Packet}
\begin{figure}[H]
    \centering
    \includegraphics[width=0.45\textwidth]{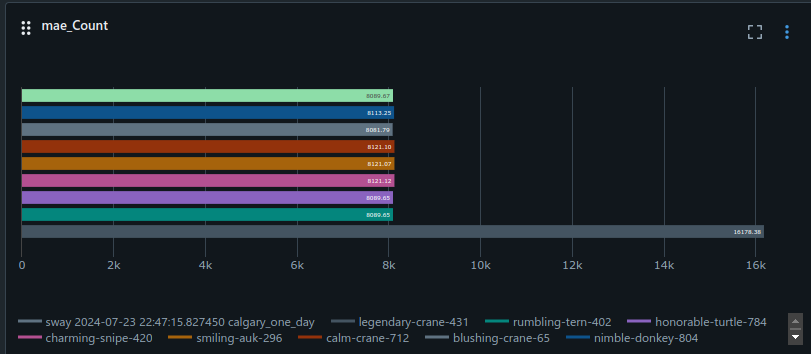}
    \caption{number of Packet}
    \label{fig:week-packetSizeDecomposition}
\end{figure}

\paragraph{sum of Packet Size}
\begin{figure}[H]
    \centering
    \includegraphics[width=0.45\textwidth]{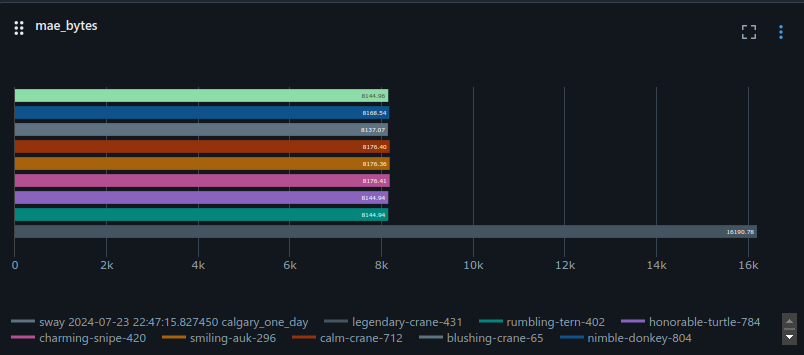}
    \caption{number of Packet}
    \label{fig:week-packetSizeDecomposition}
\end{figure}

Support Vector Regression (SVR) emerged as the most accurate model, suggesting its strong capability in handling the non-linear patterns characteristic of cloud workload data. In contrast, models like Gradient Boosting Machines (GBM) and Random Forest Regressor recorded higher errors, which may indicate issues such as overfitting or inadequacies in handling complex data structures. A detailed analysis was performed to explore the discrepancies in performance among the models. The effectiveness of SVR can be largely attributed to its ability to efficiently manage non-linear relationships within the dataset through the kernel trick. Conversely, the ensemble methods, though robust across various scenarios, did not perform optimally, potentially due to the high dimensionality and noisy characteristics of the cloud computing data.

\section{Discussion}
Notable anomalies were observed during the evaluations. For example, the unexpectedly poor performance of the Boosted Linear Regression model suggests its linear assumptions may not be suitable for the complex, variable nature of cloud workload data. This highlights the need for further investigation into feature selection and transformation to potentially enhance its predictive accuracy.

\section{Future Work}
The comparative analysis underscores the superiority of Support Vector Regression in forecasting cloud workloads within environments equipped with P4 programmable switches. Future research will focus on integrating these models into real-time systems for more refined and extensive cloud management applications.

\section{Conclusion}
The insights derived from this study are crucial for cloud service providers who aim to optimize resource allocation and enhance cost efficiency. Implementing SVR for predictive workload management could lead to significant reductions in resource mismanagement, thus improving service quality and user experience.

\bibliographystyle{unsrt}
\bibliography{lib}

\EOD

\end{document}